\documentclass[aps,prx,superscriptaddress,twocolumn]{revtex4}
\usepackage[a4paper,top=0.75in,bottom=0.75in,left=0.75in,right=0.75in]{geometry}
\usepackage{color}
\usepackage{graphicx}
\usepackage{times}
\usepackage[mathscr]{euscript}
\usepackage{verbatim}
\usepackage{amsmath}
\usepackage{amsthm}
\usepackage{amssymb}
\usepackage{amsbsy}
\usepackage{mathtools}
\usepackage{tabularx}
\usepackage{natbib}
\usepackage{bbold}
\usepackage{float}

\newcommand{\ket}[1]{| {#1} \rangle}

\DeclarePairedDelimiter\abs{\lvert}{\rvert}%
\DeclarePairedDelimiter\norm{\lVert}{\rVert}%

\begin{document}
\title{Diabatic errors in Majorana braiding with bosonic bath}
\author{Amit Nag}
\author{Jay D. Sau}
\affiliation{Condensed Matter Theory Center and Joint Quantum Institute, 
Department of Physics, 
University of Maryland, College Park, Maryland 20742-4111, USA}
\date{\today}

\begin{abstract}
Majorana mode based topological qubits are potentially subject to diabatic errors that in principle can limit the utility of topological quantum computation. Using a combination of analytical and numerical methods we study the diabatic errors in Majorana-based topological Y-junction that are coupled to a Bosonic bath in the Markovian approximation. From the study we find analytically that in the absence of a bath, the error rate can be made exponentially small in the braiding time only for completely smooth pulse shapes. Thus, pristine topological systems can reach 
exponentially small errors even for finite braid times.  
 The presence of a dominantly dissipative Markovian bath is found to eliminate this exponential scaling of error to a power-law scaling as 
$T^{-1}$ with $T$ being the braiding time. However, the inclusion of relaxation imroves this scaling slightly to go as 
 $T^{-2}$. 
 Thus, coupling of topological systems to Bosonic baths can lead to
 powerlaw in braiding time diabatic errors that might limit the speed of topologically protected operations.

\end{abstract}

\maketitle
\section{Introduction}\label{sec:intro}

\begin{figure}
\includegraphics[width=\columnwidth]{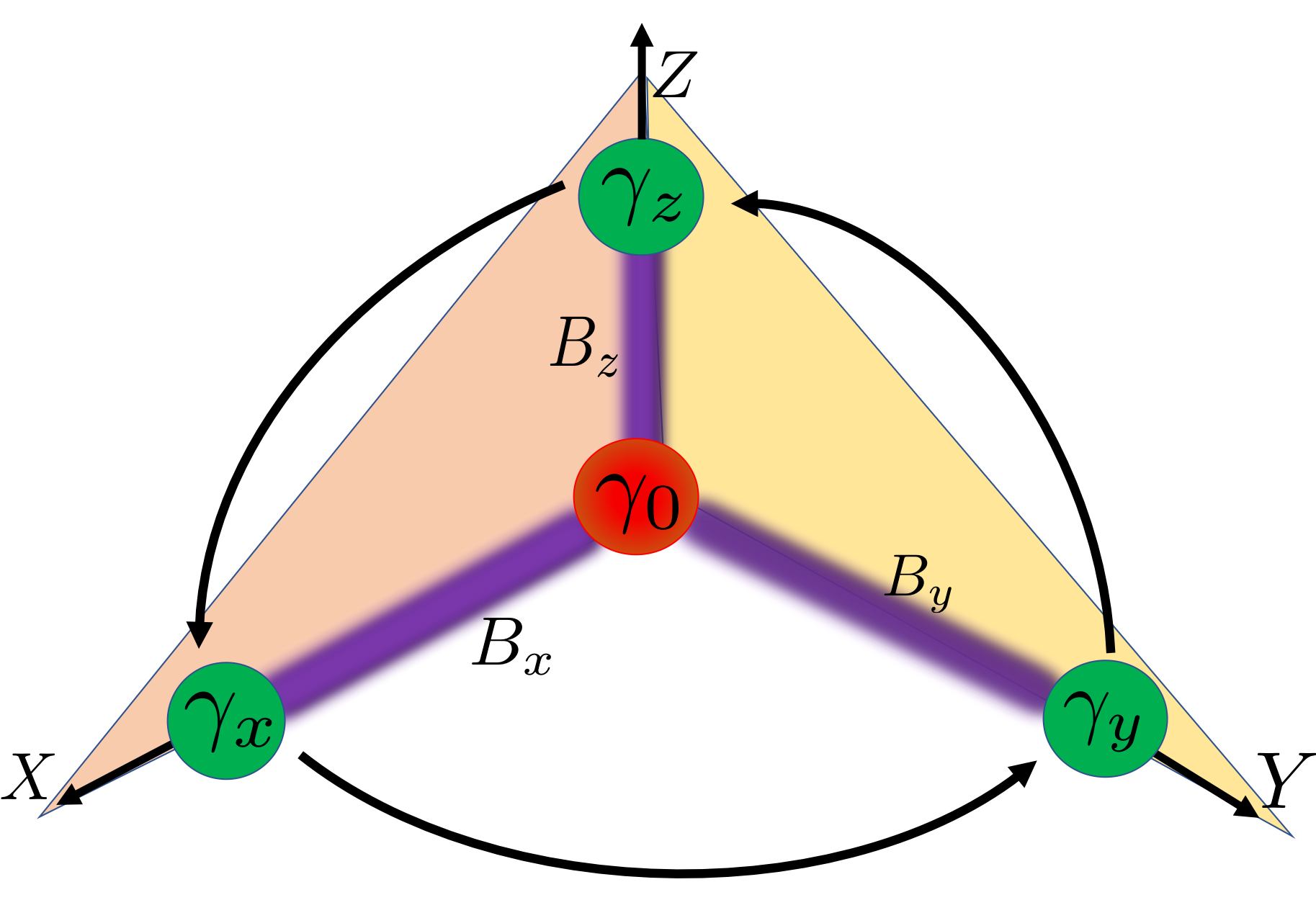}
\caption{(Color Online) Schematic description of the time-dependence of the braiding Hamiltonian(Eq.~\ref{braid-Majorana}) using the protocol in Ref.~\cite{TK}. 
The parameter $B_{\alpha}$ represents the coupling of Majoranas $\gamma_\alpha$ and $\gamma_0$. The topological degeneracy 
of the spectrum of the system is guaranteed by the requirement that at least one component of $\vec{B}=(B_x,B_y,B_z)$ vanishes.
Braiding is accomplished by successive rotations about the vanishing direction chosen to be $x,y$ and $z$.
The Bosonic bath considered here is assumed to couple as fluctuations of $B_{x,y,z}$.
}
\label{cartoon}
\end{figure}

Non-Abelian states are the building blocks of topological quantum-computation as these states carry non-locally encoded quantum information immune to decoherence~\cite{SDS_rev}. One class of such non-Abelian states are Majorana zero modes (MZMs) that emerge naturally as the ground-state of (effective) p-wave superconductors~\cite{AYK_1D,CWJB_rev}. 
MZMs (and non-Abelian states in general) appear as degenerate ground states of a stationary Hamiltonian protected by a gap wherein its simplest form involves appearance of two localized MZMs macroscopically separated in one dimensional geometry (a nanowire in an experimental setup) forming a two-fold topologically protected degenerate ground state subspace~\cite{AYK_1D,CWJB_rev}. In such systems quantum information can be encoded in the fermion parity of the degenerate ground state which depend on the occupation/non-occupation of non-local Dirac fermion mode formed out of the pair of localized MZMs. While the gap stabilizes (to exponential accuracy) the ground state subspace at finite temperature, the non-locally encoded quantum information is protected against environmental decoherence~\cite{SDS_rev,MC}. 
Concrete theoretical proposals to realize MZMs in solid-state systems~\cite{JDS_1D,YO} have led to substantial experimental efforts to realize MZMs in a laboratory. Encouraging experimental progress in this direction in the recent years~\cite{LK_1,Rokhinson,Finck,Anindya,CM2013,CM2016_1,CM2016_2,LK2016} has motivated new set of theoretical proposals that outline basic architecture of quantum gates to be realized following specific protocols to move and manipulate MZMs in order to braid and exchange a set of MZMs with each other~\cite{JA_braiding,JDS_braiding,CWJB_Braiding1,hyart2013flux}.  \\

Despite the excitement regarding the possibility of exponentially small error in topological qubits realized from MZMs, most of the studies of the protection of MZMs has been in the equilibrium phase. A less studied question is the protection of quantum information stored in MZMs to diabatic errors resulting from the finite speed of operations~\cite{MC_diabatic,TK_diabatic,Shnirman,Knapp}. Specifically, one might worry that since the topological phase of matter is a ground state property, finite gate speed might have an effect of taking the system out of the ground state in a way that introduces errors. This issue has been raised in some studies of the dynamics of braiding~\cite{Knapp} that suggest the use of a measurement based protocol as a possible way to avoid such diabatic errors. In contrast key-board based braiding protocols~\cite{JA_braiding} seem to reduce some of the diabatic errors and find error rates that scale exponentially in the rate of the process~\cite{Karzig18_1,DDV}. Another potentially critical ingredient in MZM braiding is the interaction of the MZMs with a Bosonic bath. While the effects of Bosonic bath on stationary MZMs have been studied~\cite{MC,CC,DL,FH}, its effect of diabatic braiding have been not been investigated until recently~\cite{DDV} (also see Ref.~\cite{DDV_2}) where the combined effect of (Bosonic) environmental noise and diabatic braiding is found to result in powerlaw scaling errors even for the keyboard-like braiding schemes~\cite{DDV}.

Our work focuses on the question of diabatic errors in nominally the simplest braiding protocol in a Y-junction type Majorana architecture~\cite{TK}  that is coupled to a Bosonic bath (see Fig.~\ref{cartoon}). The braiding protocol is based on the tunneling induced transport of MZMs~\cite{JDS_braiding} where the splitting between pairs of MZMs is used to exchange the decoupled MZMs $\gamma_{y,z}$ (see Fig.~\ref{cartoon}) that are 
used to store quantum information.
Fig.~\ref{cartoon} describes a particular example of a braiding protocol that leads to exchange of $\gamma_y$ and $\gamma_z$. 
Since the system is isolated (apart from the Bosonic bath) the total Fermion parity of  the system is conserved throughout the braiding protocol. 
Using the fact that at least one of the MZMs $\gamma_{x,y,z}$ are isolated from the rest of the MZMs at any time in the protocol,  
it can be shown~\cite{TK} that the two Fermion parity states remain topologically degenerate.
Both the initial and final state leave $\gamma_{y,z}$ decoupled from everything else. Therefore, the conservation of Fermion parity
equates the conservation of the MZM parity $i\gamma_y\gamma_z$, which is used to store quantum information,
to the conservation of the of the Fermion parity of the coupled pair $\gamma_{0,x}$. The latter is 
associated with excitations of the system, so that the bit-flip error is directly related to the  rate of 
exciting the system out of the ground state into the excited state. The ground and excited states are the only states in a fixed Fermion parity 
sector, so that the problem of determining the bit-flip error maps to the excitation probability of spin in a time-dependent
magnetic field.  Later we will show that the system in Fig.~\ref{cartoon} is topologically 
protected against dephasing errors in the Fermion parity basis. Therefore the problem of bit-flip errors in the system in Fig.~\ref{cartoon} 
can be mapped entirely to the relatively well-studied problem of diabatic errors for a spin in a time-dependent magnetic field (see for e.g.~\cite{childs2001robustness,yao2007restoring}).

 Despite the mapping of the system in Fig.~\ref{cartoon} to a spin in a magnetic field, the topological nature of the set-up leads to certain 
unique features when considering interactions of the system with a Bosonic bath. Microscopically, we assume 
 the Bosonic bath couples 
to the spin as a magnetic field noise similar to the classic spin-Boson model~\cite{Leggett_rev,Weiss_book}. To simplify 
our treatment, we assume that the coupling to the Bosonic bath is weak compared to temperature (much smaller than the 
 gap) so that the bath can be modeled within the Markovian approximation using the Davies prescription~ \cite{Davies,Petruccione_book}. However, unlike a 
conventional spin, the vanishing of a component of the magnetic field also implies a vanishing of the noise. This is because 
such a vanishing of a component of the magnetic field is assumed to occur because of isolation of one of the MZMs from the rest of 
the system. This leads to conservation of the associated MZM operator, which in turn leads to conservation of the associated excitation. 
Specifically, this means that for the setup in Fig.~\ref{cartoon}, the Bosonic bath is forced to decouple from the topologically protected
quantum information at the end of the process. However, this also means that any excitation generated during the dynamics of the 
effective magnetic field cannot relax away at the end of the process. This is in contrast to the dynamics in the spin-Boson model, 
where the system in contact with a zero-temperature Bosonic bath would always relax back to the ground state once the magnetic field 
becomes static at the end of the process. The absence of such relaxation leads to the a finite excitation probability in the braiding 
set-up in Fig.~\ref{cartoon}, which leads to the possibility of the bit-flip error studied in this work.

Motivated by the mapping of the set-up of Fig.~\ref{cartoon} to a spin in a magnetic field, in this work we study the probability of 
excitation of a spin in a time-dependent magnetic field that is coupled to a Bosonic bath. 
The coupling to the Bosonic bath is assumed to be small enough so that it can be studied within the Markovian 
approximation using the Davies prescription~ \cite{Davies,Petruccione_book}.  This leads to a time-dependent 
master equation which is further reduced to a Bloch equation describing spin-1/2 particles with relaxation and dephasing~\cite{Petruccione_book,JP_notes}.
To simplify the parameter space of possibilities, we assume that the temperature is low enough so that thermal excitation
 can be ignored.  We show analytically that the excitation probability in such as spin-system (corresponding to the error rate 
in Fig.~\ref{cartoon}) vanishes only polynomially as the time, $T$, within which the braid in completed.

The paper is arranged as follows. In Sec.II, we map the time dependent braiding Hamiltonian describing Fig.~\ref{cartoon} to an effective two-level spin-1/2 system coupled to a bath described by a Bloch equation. In Sec.III, following the analytical method developed in Ref.~\cite{Hagedorn}, we introduce the framework to calculate diabatic corrections to time-evolved ground-state and study diabatic corrections in the 
absence of bath coupling and present an analytical formula to calculate diabatic correction for a purely dephasing bath (i.e. in presence of bath-induced dephasing but complete absence of absence of relaxation). In Sec.IV we study our system coupled to a general bath where both dephasing and relaxation mechanisms are present. We summarize our results in Sec.V and provide details of our calculations in the appendix.

\section{Braiding Hamiltonian and bosonic bath}\label{sec:II}

In this work, we describe the system coupled to the bath by a Master equation that describes the effect of the 
bath on the time-evolution of the 
density matrix through a sequence of so-called \textit{jump} operators~\cite{JP_notes}. Below we first show how the 
Majorana braiding system shown in Fig.~\ref{cartoon} can be described by an effective spin-1/2  Hamiltonian
in a fixed Fermion parity sector. We then write down the jump operators in the Master equation that describe the 
coupling to the thermal bath.

\subsection{Spin-$1/2$ representation of braiding Hamiltonian}


The ideal system (i.e. apart from the bath) comprise four Majoranas labeled $\gamma_0,\gamma_x,\gamma_y,\gamma_z$ shown in Fig.~\ref{cartoon}. The unitary time evolution generated by the time-dependent braiding Hamiltonian over a time cycle results in exchange of two specific Majoranas. Such a braiding Hamiltonian is given by
\begin{align}
H = i\gamma_0(\vec{B}(t).\vec{\gamma}),
\label{braid-Majorana}
\end{align}
with $\vec{\gamma} = (\gamma_x,\gamma_y,\gamma_z)$. 
The Hamiltonian $H$ of the system coupled to the bath conserves Fermion parity $P=\gamma_0\gamma_x\gamma_y\gamma_z$. We use the lowest energy state of each parity $P=\pm 1$ 
to define the qubit. Choosing 
$\vec{B}$ describes coupling among Majoranas such that there is atleast one uncoupled Majorana $\gamma_i$ for $i\in\{x,y,z\}$ at any 
time ensures a topological degeneracy~\cite{TK}. 
To understand this, we label the Majorana $\gamma_i$ as the isolated Majorana at a particular time and check that  
\begin{align}
&[P,H] =[\gamma_i,H]=\{P,\gamma_i\}=0 \label{alg}.
\end{align}
If we define $\ket{\psi_{P=1}(t)}$ as the instantaneous ground state with even parity $P=1$,
it follows from Eq.~\ref{alg}  that the $\ket{\psi_{P=-1}(t)}=\gamma_i\ket{\psi_{P=1}(t)}$ 
is an eigenstate of $H$, which is degenerate with $\ket{\psi_{P=1}(t)}$ and has odd Fermion parity (i.e. $P=-1$).

\begin{figure}
\includegraphics[width=\columnwidth]{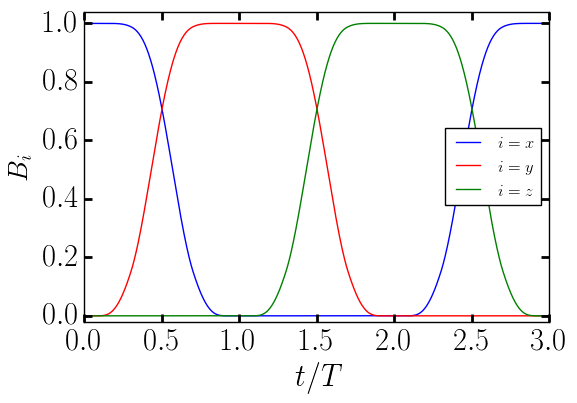}
\caption{Plot of different components of Majorana coupling field ($B_i~;~ i\in\{x,y,z\}$) appearing in Eq.~\ref{braid-Majorana} as a function of scaled-time for $s = t/T$ as described by Eq.~\ref{B(t)}}
\label{B}
\end{figure}

The time evolution of the Majoranas is governed by 
the Hamiltonian $H$ where the time-dependent magnetic field $\vec{B}(t)$ shows a three step time dependence 
shown in Fig.~\ref{B}. These steps are defined by the vanishing of one of the components $B_z, B_x$ and $B_y$, respectively, which implies the isolation of the corresponding Majorana $\gamma_{i=x,y,z}$ that is required to guarantee the topological degeneracy.
Using the relations Eq.~\ref{alg}, we can use the isolation of $\gamma_z$ (i.e. $[\gamma_z,H]=0$) at time $t=0$, to constrain the 
time-evolution of a pair of states related by:
\begin{align}
&\ket{\Psi_{ P=-1}(t)}=\gamma_z \ket{\Psi_{ P=1}(t)}\label{zcons}
\end{align} 
at time $t=0$.
Eq.~\ref{alg} lets us extend this relation from $t=0$ to the entire first segment $0<t<T$.

To determine the time-evolution for the rest of the braid we must assume that the dynamics is slow enough so that 
 the Majorana part of the system remains in the ground state. Most of the paper that follows is dedicated to determining 
the conditions for validity of this assumption. So we use this assumption for the rest of this subsection only. Using this assumption at time $t=T$, where $(\gamma_y,\gamma_0)$ are the only coupled
Majoranas, the 
 parity of the gapped wires maybe constrained as  
\begin{align}
&i\gamma_0\gamma_y\ket{\Psi_{P=1}(t=T)}=-\ket{\Psi_{P=1}(t=T)}.\label{gap}
\end{align}
Since $P=\gamma_0\gamma_x\gamma_y\gamma_z$, this also implies 
\begin{align}
&\ket{\Psi_{ P=-1}(T)}=\gamma_z\ket{\Psi_{P=1}(T)}= i\gamma_x\ket{\Psi_{P=1}(T)}.\label{ycons}
\end{align}

Repeating the arguments from Eq.~\ref{zcons},\ref{gap},\ref{ycons} two more times leads to the relation
\begin{align}
&\ket{\Psi_{ P=-1}(t)}= i\gamma_z \ket{\Psi_{ P=1}(t)} \label{braiding}
\end{align} 
at $t = 3T$. Comparing this equation to Eq.~\ref{zcons}, we realize that within the low-energy subspace spanned by $\ket{\Psi_{P=\pm 1}}$, the unitary time-evolution (ignoring an 
overall phase)
can be written as $U=e^{\pi \gamma_z \gamma_y/4}$, which is the standard representation~\cite{SDS_rev} for exchanging the Majoranas $\gamma_{z,y}$. Note that the only assumption that was crucial for this argument
was that the Majorana part of the system remains in the ground state. In principle, this allows us to consider $\ket{\Psi_P(t)}$ to be 
a wave-function of a Majorana system interacting with a Bosonic bath that we will consider later. Therefore, we can conclude that 
the qubit error (both bit-flip and dephasing) probability for the set-up in Fig.~\ref{cartoon} is given by the probability of 
excitation of the system out of the ground state in a fixed parity sector.

The flexibility of being able to focus on the excitation probability in a fixed parity sector allows us to map the Majorana dynamics 
problem to that of a spin-$1/2$ system. To see this we note that 
the Majorana operators may be expressed in terms of Pauli matrices as:
\begin{align}
\gamma_x = \sigma_x\tau_x\quad &;\quad \gamma_y = \sigma_y\tau_x \nonumber \\
\gamma_z = \sigma_z\tau_x \quad &;\quad \gamma_0 = \tau_y ,
\label{pauli}
\end{align}
(where $\tau$ and $\sigma$ are two sets of Pauli matrices) so that Eq.~\ref{braid-Majorana} can be rewritten as,
\begin{align}
H = \vec{B}(t).\vec{\sigma}\tau_z.
\label{H}
\end{align}
The Hamiltonian commutes (i.e. $[H,\hat{P}] = 0$) with the fermionic parity operator $\hat{P} = \gamma_0\gamma_x\gamma_y\gamma_z=\tau_z$. 
Within the $P=\tau_z=1$ parity sector, the dynamics of the system may be described by the reduced \textit{2-level} Hamiltonian,
\begin{align}
H_{2Level} = \vec{B}(t).\vec{\sigma}.
\label{2Level}
\end{align}

 Derivative discontinuities in the time-dependence of $\vec{B}(t)$ are known to introduce
diabatic errors in Majorana qubits that scale as a power-law in $T$~\cite{Knapp}. To avoid such diabatic errors the profile for $\vec{B}(t)$ in Fig.~\ref{B} 
is chosen to be a piece-wise continuous functions where all derivatives are continuous (i.e. $C^{\infty}$). The specific form in Fig.~\ref{B} 
that accomplishes this level of smoothness is given by 
\begin{align}
B_x &= 
\begin{cases}
    \cos(\theta(s)) ,& 0\leq s< 1\\
    0 ,& 1\leq s<2               \\
    \sin(\theta(s-2))  ,& 2\leq s\leq 3
\end{cases} \nonumber \\
B_y &= 
\begin{cases}
    \sin(\theta(s)) ,& 0\leq s< 1\\
    \cos(\theta(s-1))  ,& 1\leq s <2           \\
    0 ,& 2\leq s\leq 3
\end{cases} \nonumber \\
B_z &= 
\begin{cases}
    0 ,& 0\leq s< 1\\
    \sin(\theta(s-1)) ,& 1\leq s <2               \\
    \cos(\theta(s-2))  ,& 2\leq s\leq 3 .
\end{cases} 
\label{B(t)}
\end{align}
where 
\begin{align}
\theta(s) = \frac{\pi\int_0^s ds' e^{-1/s'(1-s')}}{2\int_0^1 ds' e^{-1/s'(1-s')}}, \quad s\in[0,1]
\label{theta}
\end{align}
and 
\begin{align}
s\equiv t/T .
\label{scaled}
\end{align} 
 is the dimensionless time parameter.

\subsection{Master Equation}
Here we describe the Master equation that we use to model the interaction of the system with a parity-conserving thermal bath
 within the Born-Markov approximation. As discussed such a bath can only generate errors by flipping the system 
to the excited state of the spin Hamiltonian $H_{2level}$ in Eq.~\ref{2Level}. Therefore, similar to the previous subsection, it 
suffices us to derive a Master equation for the spin-1/2 system.

The total Hamiltonian describing the system interacting with a bosonic bath is,
\begin{align}
H_T = H_{2level} + H_{SB} + H_{B},
\end{align}
where, $H_{2level}, H_B, H_{SB}$ are system Hamiltonian, bosonic-bath Hamiltonian and Hamiltonian describing system-bath coupling, respectively. In general, the system-bath Hamiltonian have the following form,
\begin{align}
H_{SB} = \sum_k \hat{A}_k \hat{\Gamma}_k,
\label{system-bath}
\end{align}
where $\hat{A}_k$ and $\hat{\Gamma}_k$ are system and bath operators, respectively, satisfying $\hat{A}_k^\dagger = \hat{A}_k$ and $\hat{\Gamma}_k^\dagger = \hat{\Gamma}_k$. 

Following the prescription introduced by Davies~\cite{Davies}, under the assumptions of weak system-bath coupling (\textit{Born} approximation) and memory-less bath (\textit{Markov} approximation: bath correlation time vanishes on system time scales), the master equation describing the time evolution of density matrix of the system can be expressed in the \textit{Lindblad} form~\cite{Davies,Petruccione_book} as 
\begin{align}
\dot{\rho}_S(t) = -i[H_{2level} , \rho_S (t)] + \mathcal{D}(\rho_S(t)),
\label{Lindblad}
\end{align}
where the bath term is written as:
\begin{align}
\mathcal{D}(\rho_S(t)) &= \sum_{\nu,i} J_i^{\nu}(t)\rho(t) J_i^{\nu\dagger}(t)\eta_k(\nu)\nonumber \\
 &-\frac{1}{2}\eta_k(\nu)\{ {J_i^{\nu}}(t)^\dagger J_i^{\nu}(t),\rho_S(t)\}.
\label{Lindblad1}
\end{align}
The above bath-induced dissipation term is written in terms of projections $J_k^\nu(t)$ 
of the 
system operators $\hat{A}_k(t)$, referred to as "jump operators" that are  written as 
\begin{align}
J_k ^\nu (t)= \sum_{e ' - e = \nu} 
\Pi(e)\hat{A}_k(t)\Pi(e '),
\label{jump}
\end{align}
with $\Pi(e)$ being the projection operator to an eigenspace spanned by eigenstates of $H_{2level}$ with eigenvalue $e$. 
The coefficients $\eta_k(\nu)$ in Eq.~\ref{Lindblad1} are correlators of the bath operators that are written as 
\begin{align}
&\eta_{k}(\nu)  = \mathrm{Re}\int_{-\infty}^\infty ds e^{-i\nu s}\left\langle \hat{\Gamma}_{k}^\dagger (s)\hat{\Gamma}_k (0)\right\rangle
\end{align}
where, $\langle\dotsb\rangle$ denotes thermal average, $\mathrm{tr}_B(\dotsb\rho_B)$. 
Depending on the value of $\nu$, the jump operators $J_k^\nu$ are classified as excitations ($\nu>0$), relaxation ($\nu<0$) 
or dephasing ($\nu=0$).

The bath operators couple to each of the components of $B_i$ (where $i=\{x,y,z\}$) in  a way where the bath coupling to a 
particular wire in Fig.~\ref{cartoon} vanishes when the corresponding $B_i$ vanishes. This is essential to represent the fact that the 
vanishing $B_i$ results from decoupling of $\gamma_i$ from $\gamma_0$. To satisfy this condition, we choose the system 
operators for the bath coupling as 
\begin{align}
&\hat{A}_i = s_iB_i\sigma_i.
\end{align}
Using Eq.~\ref{jump}, the \textit{jump operators} corresponding to the above choice of $\hat{A}$ operators are expressed as,
\begin{align}
J_i^{\nu} (t) &= s_i a_i(t) B_i(t) |1(t)\rangle\langle 0(t)|; \quad J_i^{-\nu} = J_i^{\nu\dagger}\nonumber\\
 J_i^{\nu=0} &= s_i B_i(t) (a^0_i(t)  |0(t)\rangle\langle 0(t)| +a^1_i(t)  |1(t)\rangle\langle 1(t)|)
\label{jump-op}
\end{align}
where,
\begin{align}
a_i(t) &= \langle 1(t)|\sigma_i|0(t)\rangle \nonumber \\
a^0_i(t) &= \langle 0(t)|\sigma_i|0(t)\rangle \nonumber \\
a^1_i(t) &= \langle 1(t)|\sigma_i|1(t)\rangle,
\label{jump_matrix}
\end{align}
$s_i$ being the time-independent system-bath coupling strength with $i\in\{x,y,z\}$ and $|0(t)\rangle$ and $|1(t)\rangle$ denoting the instantaneous ground and the excited states of $H_{2level}(t)$, respectively.

We wish to focus on only those bath-effects that arise due to time-dependence of the Hamiltonian. In other words we want to explicitly avoid  any bath-induced effect that do not vanish in the adiabatic limit provided the system is initialized in the ground state. The only such environmental effect is thermal excitation associated with $J^{\nu = 2}$ (energy gap above ground state is 2 for $\norm{\vec{B}}=1$) operator. We explicitly set temperature to zero to completely suppress these thermal excitations or equivalently,
\begin{align}
\eta(\nu = 2) &= 0. \nonumber \\
\end{align}

Since the braiding process involves sequence of three identical clockwise  $\pi/2$ rotations of $\vec{B}$ along $\hat{z},\hat{x}$ and $\hat{y}$ axes respectively, as shown in Fig.~\ref{cartoon}, we focus on the first sequence where $\vec{B}$ is restricted to XY plane without loss of generality, setting $B_z = 0$.  Influence of the bath is captured by the strength of the system-bath couplings (captured by $s_x$ and $s_y$), the relaxation strength governed by $\eta(\nu = -2)$ (since the gap in the system is 2) and dephasing strength governed by $\eta(\nu = 0)$.
Since the time-dependence of the Hamiltonian does not affect the gap in the system $\eta(\nu = -2)$ and $\eta(\nu = 0)$ are fixed parameters determined by entirely by the microscopic properties of the bath Hamiltonian. From here onwards, for the sake of brevity, we relabel the relaxation strength, $\eta(\nu = -2)$ and the dephasing strength, $\eta(\nu = 0)$ with new symbols, $\eta(\nu = -2) \equiv \eta$ and $\eta(\nu = 0) \equiv \eta_0$.

\subsection{Bloch equation}

The density matrix $\rho_S(t)$ appearing in the master equation (Eq.~\ref{Lindblad} can be parametrized in 
terms of a Bloch vector $R\equiv(r_x,r_y,r_z)$ defined by
\begin{align}
\rho_S(t) &= \frac{1}{2}(r_x(t)\sigma_x + r_y(t)\sigma_y + r_z(t)\sigma_z + \mathbb{1}).
\end{align}
Rewriting the Master equation (Eq.~\ref{Lindblad}) in terms of $\vec{R}$ leads to the so-called Bloch equation 
\begin{align}
\epsilon\dot{\vec{R}} = 2[\vec{B}\times\vec{R}  + (\alpha -\beta)\vec{B}\times(\vec{B}\times\vec{R}) - 2\beta(\vec{B}+\vec{R})] 
\label{vecBloch}
\end{align}
where,
\begin{align}
\beta(s) &= \frac{1}{4}F(s)\eta;\quad \text{(effective~relaxation)}\nonumber \\
\alpha(s) &= F^0(s)\eta_{0};\quad \text{(effective~dephasing)}
\label{alphabetadef}
\end{align}
with,
\begin{align}
F &= s_x^2B^2_y(s)B_x^2(s) + s_y^2B^2_x(s)B_y^2(s) \nonumber \\
F^0 &= s_x^2B_x^4(s) + s_y^2B_y^4(s)
\label{F}
\end{align}  
and $\epsilon = 1/T$. Note that the time derivatives in the above equation refer 
to the rescaled time $s=t/T$.

The Bloch equation can be more compactly expressed in a matrix form as,
\begin{align}
\epsilon\frac{d}{ds}{R} = MR + 4\beta (R_0 - R)
\label{wnoise}
\end{align}
with, $R_0(s) \equiv -\vec{B}(s)$ being a null vector of $M$,  $M = 2(A + S)$ and $S = (\alpha - \beta)A^2$ 
where,
\begin{align}
A &= 
\begin{pmatrix}
0 & 0 & B_y \\
0 & 0 & -B_x \\
-B_y & B_x & 0 
\end{pmatrix}
\label{BlochDef}
\end{align}

\section{Adiabatic expansion for Bloch Equation}\label{sec:III}
As discussed in Sec.~\ref{sec:intro}, the diabatic error for the Majorana qubit in Fig.~\ref{cartoon} is related to the probability of transition out of the ground state.
Such a transition can occur in any one of the three steps of the braiding protocol, so that the order of magnitude of the error can be 
estimated from the transition probability in any one step. 
We consider the first step of the braiding process where $B_z=0$ and calculate the diabatic error incurred in the process which involves rotating $\vec{B}$ along $\hat{z}$ axis by $\pi/2$ in the clockwise direction from the initial orientation along $\hat{x}$ to final orientation along $\hat{y}$.  
The system is initialized in the ground state of $H_{2Level}$ to which the corresponding Bloch vector is $R(s = 0) = -\vec{B}(s = 0) = (-1,0,0)$. 
Given this initial condition, the solution of the Bloch equation (i.e. Eq.~\ref{wnoise}) in the extreme adiabatic limit (i.e.  $\epsilon=0$), is written as:
\begin{align}
R(s) = R_0(s). 
\end{align} 
At non-zero $\epsilon$, the time-dependent solution $R(s)$ of the Bloch equation~\ref{wnoise},  referred to as \textit{time-evolved Bloch zero-vector} (TBZV),
differs from the adiabatic limit $R_0(s)$. The magnitude of the difference at $s=1$ 
\begin{align}
\mathcal{E} \equiv \norm{R(1)-R_0(1)},
\end{align}
quantifies the diabatic error for the braiding.

Let us now consider the solution of the Bloch equation Eq.~\ref{wnoise} for infinitesimal $\epsilon$. We first focus on the case of vanishing 
relaxation (i.e. $\eta=\beta=0$) where the Bloch equation reduces to 
\begin{align}
\epsilon\dot{R} = M R.
\label{BEA}
\end{align}
The finite relaxation situation will be dealt with in Sec.~\ref{sec:IV} using a slightly different technique.
The solution of the Bloch equation in the above form admits an adiabatic series expansion written as:
\begin{align}
R(s) = R_0(s) + \epsilon R_1(s) + \epsilon^2 R_2(s)\dotso
\label{expansion}
\end{align}

Substituting the above ansatz in Eq.~\ref{BEA} and equating both sides of the equation at each order in $\epsilon$ we formally solve for $j^{th}$ order correction to $R_0(s)$ for $R(s)$ (see App.~\ref{app:B} for details),
\begin{align}
R_j &= f_{j-1}R_0 + M^{-1}\dot{R}_{j-1}
\label{recursive}
\end{align}
with $f_{j-1}(s)$ given by,
\begin{align}
f_{j-1}(s) = \int_0^s ds~\dot{R}_0^TM^{-1}\dot{R}_{j-1}.
\label{f_j}
\end{align}
The matrix $M$ is singular so that it is crucial to note that Eq.~\ref{BEA} implies that $R_0^T\dot{R}\propto R_0^T M R=0$. 
Additionally this implies that the generalized inverse $M^{-1}$ can be defined so that $M^{-1}\dot{R}$ is orthogonal to $R_0(s)$.
Motivated by this, it is convenient to split the expansion (Eq. ~\ref{expansion}) into two parts 
\begin{align}
R(s) &= f(s) R_0(s) + R_{\perp}(s)
\label{||perp}
\end{align}
where, $R_0^T(s)R_\perp(s)=0$ and $f(s)$ and $R_{\perp}(s)$ are expanded as:
\begin{align}
f(s) &= 1 + \epsilon f_0(s) + \epsilon^2f_1(s) +\dotso \nonumber \\
R_{\perp}(s) &= M^{-1} (\epsilon \dot{R}_0(s) + \epsilon^2 \dot{R}_1(s)+ \dotso).
 \label{fs}
\end{align}
Taking repeated derivatives of Eq.~\ref{recursive}(see App.~\ref{app:B} for details) we can show that 
$\dot{R}_j(s=1)$ vanishes when all derivatives of $R_0$ vanish, $R^{(n)}_0(s) \equiv -\vec{B}^{(n)}(s) = 0$ at $s=0,1$. 
 This implies that $R_\perp(s=1)$ vanishes as well only when the magnetic field $\vec{B}(s)$ goes to 0 smoothly at $s=0,1$,
which is a condition that is ensured by the specific time-dependence of the magnetic field chosen (i.e. Eq.~\ref{B(t)}).
More precisely, considering the expansion for $R_\perp$ (Eq.~\ref{fs}) to a finite order $n$, one can show that 
\begin{align}
 \norm*{R_{\perp}(s=1)}/\epsilon^n &= 0~as~\epsilon\rightarrow 0  
\label{eq:estimatem1}
\end{align}
for every integer $n$. This shows that $R_\perp(s=1)$ vanishes faster than any power-law in the adiabatic limit. 
A numerical analysis shows that 
\begin{align}
&\norm*{R_\perp(s=1)}\sim e^{-\sqrt{T}},
\label{eq:estimate0}
\end{align}
which is similar to (but slower than)  exponential scaling with $T$.

\begin{figure}
\includegraphics[width=\columnwidth]{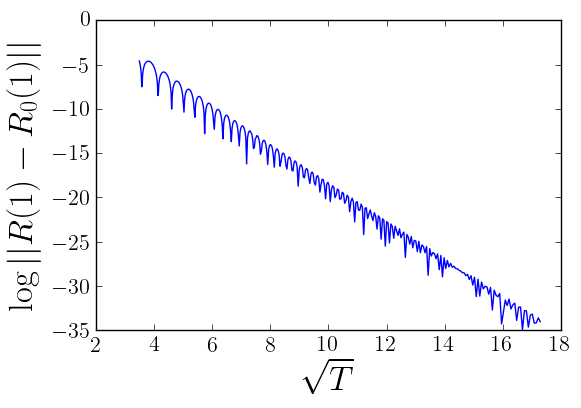}
\caption{Numerical plot of log of norm of difference between time evolved and instantaneous Bloch zero-vector at scaled time $s=1$ as a function of $\sqrt{T}$ for strength $s_x = s_y=0$ obtained using Eq.~\ref{wnoise}.  The system is initialized in the ground state of $H_{2Level}$ or equivalently the initial value of $R(s=0)$ in the Bloch equation is set as $R_0(0)=-\vec{B}(0)$. The plot shows exponential decay in $\mathcal{E}$ as a function of total time $T$ when system-bath coupling is zero consistent with the result derived in App.~\ref{app:B}. Specifically,  $\mathcal{E} = \norm{R(1)-R_0(1)}\sim e^{-\sqrt{T}}$.}
\label{noise-free}
\end{figure}

In the absence of any bath (i.e. $\eta = \eta_0 =\alpha=\beta=0$) the dynamics of $R(s)$ is unitary and $\norm*{R(s)}=1$.
Thus the bound on $R_\perp(s=1)$ (Eq.~\ref{eq:estimate0}) also implies a bound on the diabatic error for the isolated qubit given by 
\begin{align}
\mathcal{E} = \norm{R(1)-R_0(1)}\sim e^{-\sqrt{T}}.
\label{exponential}
\end{align} This is completely consistent with the numerical results in Fig.~\ref{noise-free} and also with those derived in Ref.~\cite{Hagedorn}, which suggests an exponential-like dependence of diabatic error $\mathcal{E}$ on $T$.

In the presence of dephasing $\eta\neq 0$, the unitarity constraint  $\norm*{R(1)}=\norm*{R_0(1)}$ does not apply, but one can use the 
essential vanishing of $R_\perp(s=1)$ (Eq.~\ref{eq:estimatem1}) in combination with Eq.~\ref{||perp}
to write 
 \begin{align}
 R(1) =  f(1) R_0(1).
 \label{R1A}
 \end{align}
Evaluating the recursive relation Eq.~\ref{f_j}, one finds (see  App.~\ref{app:B} for details) a non-vanishing $O(\epsilon)$ contribution to $f(1)$ that is 
written as  
\begin{align}
 f_0(1) = -\int_0^1 \frac{\alpha\omega^2}{2(1+\alpha^2)}. 
 \label{F0A}
\end{align}
 This leads to a power-law diabatic error estimate, 
 \begin{align}
\mathcal{E} &= \norm{R_0(1)}\bigl|( 1-f(1))\bigr| \approx -\epsilon  f_0(1)
\label{EBA}
\end{align}
Note that such error term (as well as the other powers of $\epsilon$) dominate over $e^{-\sqrt{T}}$ estimate for the $R_\perp(s=1)$ component
that was ignored but would vanish in the absence of a bath ($\alpha =0$) restoring the exponential estimate for the isolated 
system situation.

\begin{figure}
\includegraphics[width=\columnwidth]{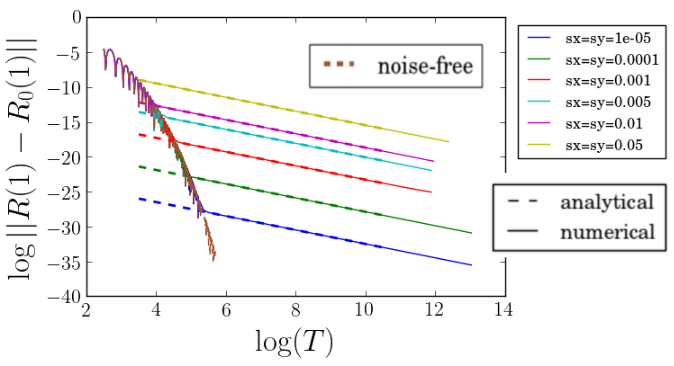}
\caption{Plot of $\mathcal{E}$ as a function of time $T$ for different values of effective dephasing strength $\alpha$ characterized by values of $s_x = s_y$ and absence of effective relaxation, $\beta = 0$ (see Eq.~\ref{alphabetadef}).  Relaxation and dephasing parameters, $\eta$ and  $\eta_{0}$, that solely depend on bath Hamiltonian are both set to $\eta=\eta_{0}=1$. The solid line is obtained by solving the Bloch equation (Eq.~\ref{BEA}) and is compared against (dashed curve) theoretical bound given by Eq.~\ref{EBA},\ref{F0A}, i.e., dashed curve is calculated value of $\log[ (1/T)f_0(1)]$ using the expression given in Eq.~\ref{F0A}. The low-T region characterized by the oscillatory section of the curves are independent of system-bath coupling strength and follows closely the numerical values obtained by solving the Bloch equation (Eq.~\ref{wnoise}) in absence of bath denoted by the dashed-brown-colored curve.}
\label{beta0}
\end{figure}

In Fig.~\ref{beta0}, we plot $\mathrm{log}\mathcal{E}$ as a function of $\log(T)$ for different system-bath coupling strengths and compare it against numerical curve obtained by solving Eq.~\ref{BEA}.  We find that numerically exact calculation of $\mathcal{E}$  validates the asymptotic (large $T$) limit of the diabatic error (Eq.~\ref{EBA}) obtained for truncation at first-order in epsilon.
We find that the scaling region that follows Eq.~\ref{EBA} begins rather abruptly above a value of $T$ that depends on the 
dissipation strength. The behavior below this threshold coincides with the behavior of the dissipationless system shown in Fig.~\ref{noise-free}.
Thus, the strength of the dissipation determines the critical braiding rate above which the exponential behavior of diabatic error (Eq.~\ref{exponential}) 
crosses over into power-law error described by Eq.~\ref{EBA}.

\section{General system-bath coupling} \label{sec:IV}
We now consider the error $\mathcal{E}$ in the presence of finite relaxation. In order to solve for $R$ in Eq.~\ref{vecBloch} for the case of finite $\beta$ we switch to a new vector-variable $\vec{\Pi}$,
\begin{align}
R(s) = U_0(s)(-\hat{x}+\vec{\Pi} (s))
\label{defPi}
\end{align}
where, $U_0(s)$ is the rotation matrix such that $R_0(s) = U_0R_0(0)$ ($R_0(0) = -\hat{x}$ being the initial condition on $R$) given by
\begin{align}
U_0 &= \begin{pmatrix}
\cos\theta& \sin\theta &0 \\
-\sin\theta & \cos\theta &0 \\
0 & 0 &1
\end{pmatrix}.
\label{basic}
\end{align}  
From Eq.~\ref{defPi} it follows, the initial condition on $\Pi (s)$  is $\Pi(0) = (0,0,0)$. Since $R_0(s)$ is the solution to the Bloch equation in $\epsilon\rightarrow 0$ limit, ${\Pi} (s)$ must have perturbatively small norm in $\epsilon$.       

The equation of motion for $\Pi (s)$ is found to be,
\begin{align}
\epsilon\left(\dot{\vec{\Pi}} +\dot{\theta}\hat{z}\times\vec{\Pi} - \dot{\theta}\hat{y}\right)&= 2\hat{x} \times\vec{\Pi}\nonumber \\
&+ 2(\alpha - \beta)\hat{x}\times(\hat{x}\times\vec{\Pi}) - 4\beta \vec{\Pi}.
\label{Pi}
\end{align}
On the right hand side of the equation, the operator acting on $\Pi$ can be diagonalized along the basis unit vectors $\{\hat{x},\hat{j}_{+},\hat{j}_{-}\}$ with $\{-4\beta,2\lambda_{+},2\lambda_{-}\}$ being the corresponding eigenvalues where we have defined, 
 \begin{align}
  \hat{j}_{\pm} &= \frac{1}{\sqrt{2}}(\hat{y} \pm i\hat{z})  \\
  \lambda_{\pm} &= \mp i-(\alpha + \beta).
  \end{align}
  Representing $\vec{\Pi}$ in this basis set,
 \begin{align}
 \vec{\Pi} = \pi_x\hat{x} + \pi_+\hat{j}_{+} + \pi_-\hat{j}_{-}
 \end{align}
 leads to the following coupled-differential-equation,
\begin{subequations}
  \label{foo}
  \begin{gather}
    \epsilon\left[\dot{\pi}_x - \frac{\dot{\theta}}{\sqrt{2}}(\pi_+ + \pi_-)\right] = -4\beta\pi_x  \label{pix}  \\
    \epsilon\left[\dot{\pi}_\pm + \frac{\dot{\theta}}{\sqrt{2}}\pi_x - \frac{\dot{\theta}}{\sqrt{2}}\right] = 2\lambda_\pm\pi_\pm  
   \end{gather}
\end{subequations}

Finite relaxation $\beta> 0$ ensures that at sufficiently long time $T$ (or sufficiently small $\epsilon$), the memory of the 
initial conditions at $s=0$ are exponentially suppressed. In this case, Eq.~\ref{pix} suggests  $\pi_x(0<s<1)$ is $O(\epsilon)$
so that, to lowest order in $\epsilon$,
  \begin{align}
 \pi_\pm \approx -\frac{\dot{\theta}\epsilon}{\sqrt{2}\lambda_\pm}.
 \label{pi+sol}
  \end{align}
 This implies $\pi_+(1)=\pi_-(1)\approx0$ as $\dot{\theta}|_{s=1}=0$ (see Eq.~\ref{theta}). Therefore, we conclude
 $\mathcal{E} = \norm{R(1)-R_0(1)}\approx \pi_x(1)$. Substituting $\pi_\pm$ expressions in Eq.~\ref{pix}, it can be solved for $\pi_x$ resulting in 
 \begin{align}
\mathcal{E} \simeq  \pi_x(1) = -\epsilon\int_0^1 \dot{f}_0e^{-\frac{4}{\epsilon}\int_s^1\beta ds'}ds. 
\label{fullTh2}
 \end{align}
 where we have defined,
\begin{align}
{f}_0(s) &= -\int_0^s \frac{(\alpha+ \beta)\dot{\theta}^2}{2(1+(\alpha + \beta)^2)}. 
\label{fOrder1}
\end{align}
The behavior of $\mathcal{E}$ in the large-T limit is analyzed using saddle-point method detailed in the App.~\ref{app:E} to obtain the following asymptotic form for the diabatic error,
\begin{align}
\mathcal{E} \sim e^4T^{-2}.
\label{asymptotic}
\end{align}

Note that Eq.~\ref{fullTh2} reduces to Eq.~\ref{EBA} in absence of relaxation, i.e. $\beta\rightarrow0$, showing that 
the approach in this section is not inconsistent with that in the previous section for the case without relaxation.
An examination of the exponential factor in  Eq.~\ref{fullTh2} tells us that for the case where relaxation is weak compared to dissipation (i.e. $\beta\lesssim \alpha$), we should expect an intermediate regime of $\epsilon$ where $\beta$ might still be ignored so that the 
error would scale as $\mathcal{E}(T)\sim T$ as concluded in the previous section.
 
\begin{figure}
\includegraphics[width=\columnwidth,height=0.7\columnwidth]{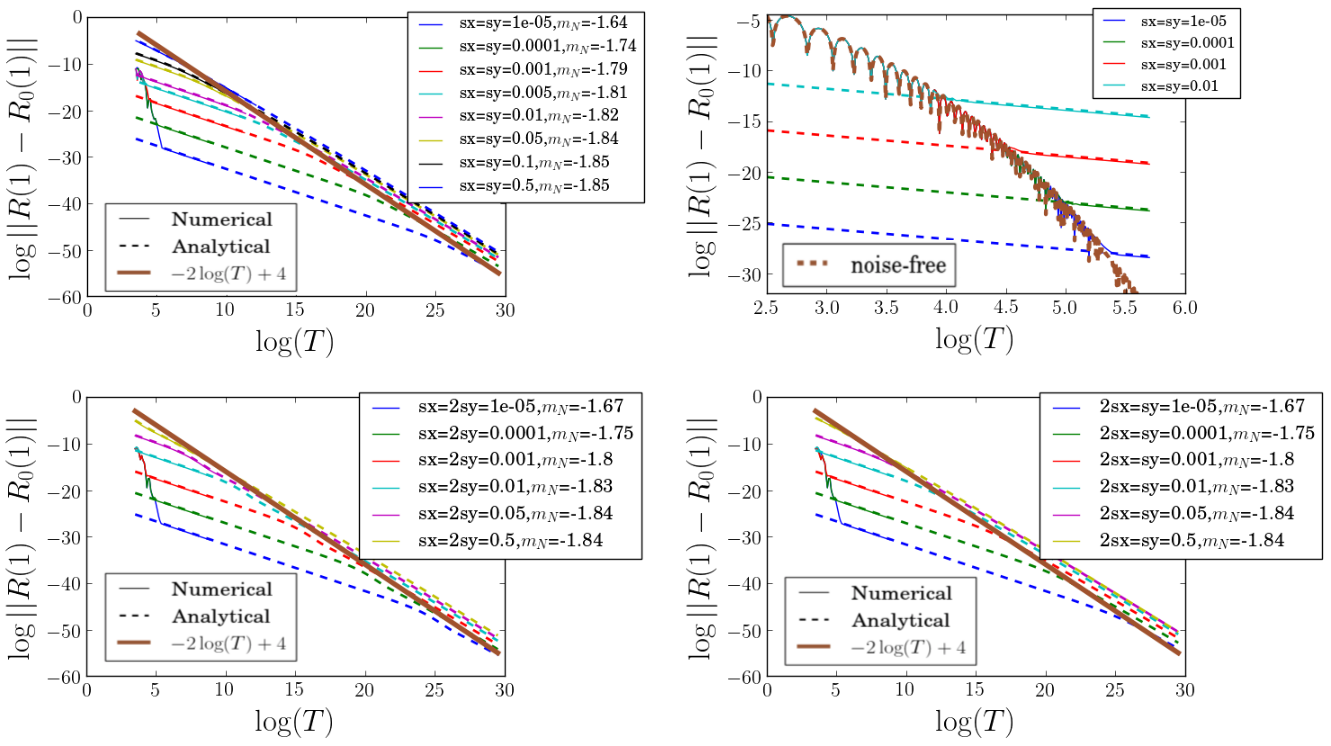}
\caption{(Top left) Plot of $\mathcal{E}$ as a function of total-time taken by the diabatic drive $T$ for different values of effective dephasing, $\alpha$ and relaxation strength, $\beta$ characterized by the shown values of $s_x = s_y$ (see Eq.~\ref{wnoise}-\ref{F}). Relaxation and dephasing parameters, $\eta$ and  $\eta_{0}$, that solely depend on bath Hamiltonian are both set to $\eta=\eta_{0}=1$. for all panels in this figure. For each set of parameters, the analytical curve (dashed line) is obtained by plotting the expression in Eq.~\ref{fullTh2} with $f_0$ given by Eq.~\ref{fullTh11}. The numerical curve (solid line) is obtained by numerically solving the Bloch equation in Eq.~\ref{wnoise}. The slope $m_N$ (given in legend) is calculated using the data in the neighborhood of log(T) = 29.5 on the X axis for each curve. The asymptotic dependence on $T$ is given by $e^4T^{-2}$ (Eq.~\ref{asymptotic}), represented in the figure by the solid (thick) brown curve. (Top right) Magnified view of the top left panel in the small-T-region.  The values obtained for the special case of zero dephasing and relaxation is plotted using (thick) brown dashed curve. For each coupling strength and sufficiently small values of T, the $\mathcal{E}=\log (R(1)-R_0(1))$ behavior is independent of system bath coupling strength and is well captured by the exponential dependence on T (see Fig.\ref{noise-free}). The abrupt change from the exponential to a polynomial dependence on T occurs when the analytic estimate of  $\mathcal{E}$ coincides with its value under zero system-bath coupling assumption. (Bottom panels) Similar to the top panels, log norm of difference between time evolved and instantaneous Bloch vector at scaled time $s=1$ ($t/T\mod{3}=1$)as a function of log of total time $T$. However, unlike the top panels $s_x=2s_y$ and $s_y=2s_x$ for bottom left and right panels respectively. The qualitative behavior is similar to ones observed in the top panels.}
\label{alphabeta}
\end{figure}

Therefore as $T$ is increased we expect the error rate 
$\mathcal{E}$ to crossover from the isolated system limit (Eq.~\ref{exponential}) to the dominantly dissipative system(Eq.~\ref{EBA}) to the asymptotics with 
relaxation (Eq.~\ref{asymptotic}).
  The numerical  plot of the error rate $\mathcal{E}$ in the general case, Fig.~\ref{alphabeta} indeed shows this 
expected pattern of crossovers.
These expectations turn out to be quite accurate as seen from the numerical simulations in Fig.~\ref{alphabeta}, which shows the numerical estimate of $\mathcal{E}$ vs total time $T$ in log-log scale for dephasing strength $\eta_0$, and relaxation strength $\eta$, both set equal to 1. It is interesting that the crossover behavior expected for the limit $\eta_0\gg \eta$ seems to apply even to $\eta_0=\eta=1$ chosen for Fig.~\ref{alphabeta}. Finally, we see in Fig.~\ref{alphabeta} that as we decrease the bath coupling strength $s_x\sim s_y$, the crossover scale modes to 
higher $T$ as expected. While Eq.~\ref{fOrder1} predicts the correct crossover behavior, it turns out to be off by a $T$-independent 
scale factor in the pre-asymptotic dissipation dominated regime.
We have remedied this by adapting the analysis in Sec.~\ref{sec:III} to finite (but small $\eta$) as described in App.~\ref{app:E}. This analysis leads to a somewhat different approximation for $f_0(s)$ which is written as
\begin{align}
\label{fullTh11}
f_0(s) &= -\int_0^s \frac{(\alpha- \beta)\dot{\theta}^2}{2(1+(\alpha - \beta)^2)}. 
\end{align}
This more exact theoretical form, which is the analytic approximation used in the dashed lines 
in Fig.~\ref{alphabeta}, can be seen from the figure to fit the numerical results very well in both the dissipative 
and relaxation dominated regimes.

\section{Summary}
In this work we have studied the diabatic error rate $\mathcal{E}(T)$ as a function of braiding time $T$ (in units of 
inverse gap) of the Y-junction braiding protocol (shown in Fig.~\ref{cartoon}) by mapping it 
to the problem of excitation probability of a spin in a time-dependent magnetic field. 
 Consistent with previous work~\cite{Hagedorn} we find (see Eq.~\ref{exponential}) that one can reduce diabatic errors~\cite{Knapp} 
in isolated Majorana systems to be
exponentially (scaling as $\mathcal{E}(T)\sim e^{-\sqrt{T}}$) small in the braiding time $T$ (in units of inverse gap)  by choosing the time-dependence of the Hamiltonian
 to be completely smooth (including the beginning and end of the protocol).
 This analytic result explains the recent obeservation of exponentially suppressed 
diabatic errors apparent from numerical simulation of some braiding protocols~\cite{Karzig18_1} 
In fact, while the requirement of a completely smooth time-dependent Hamiltonian of the form of Eq.~\ref{H} (also see Eq.~\ref{B(t)}) seems 
fine-tuned, it is quite natural in a topological system where MZM splitting is tuned either by chemical potential~\cite{JA_braiding,sau2010universal,Karzig18_1} or by screening charging energy through tunable Josephson junctions~\cite{hyart2013flux,clarke2016flux}. In both these cases, the tunneling is exponentially suppressed as one tunes the Majorana wire deep 
into the topological phase or introduces a strong Josephson coupling between the Majorana wire and a bulk superconductor. 
The main focus of our work is to consider the effect of interaction of the topological system with a
 Fermion parity conserving Bosonic bath, such as phonons or plasmons 
on the diabatic error rate for braiding. Similar to previous work~\cite{DDV}, we assume the bath to be weakly 
coupled so that we can treat the bath within the Markovian approximation. 
We describe the dynamics of this system within the Markovian approximation  by a Bloch equation. 
We find that the coupling to such a Bosonic bath generically changes the asymptotic of the diabatic error from 
exponential in the braiding time $T$ to power law (i.e. error scaling as $\mathcal{E}(T)\sim e^4T^{-2}$ (see Eq.~\ref{asymptotic}). 
as expected, the general result including relaxation leads to much lower excited state probability than in the absence of 
relaxation. We find from a controlled analytic solution that the error rates in the presence of dephasing from the bath but 
no relaxation decreases the slowest scaling as $\mathcal{E}(T)\sim T^{-1}$.  

In addition to the asymptotic forms we study the dependence of the 
diabatic error rate on the system-bath coupling strength through direct numerical simulations. For purely dephasing 
bath (i.e. no relaxation), we find (see Fig.~\ref{beta0}) that the dephasing strength determines the cross-over from 
the $e^{-\sqrt{T}}$ scaling error rate that is expected in the absence of a bath to the $T^{-1}$ scaling of the error rate
expected for a purely dephasing bath. Increasing the dephasing strength leads to the crossover occuring at 
shorter braiding time $T$, in turn leading to error rates that increase with increasing dephasing. As seen from our 
results in Fig.~\ref{alphabeta}, adding relaxation in addition to dephasing leads to an additional 
relaxation strength dependent cross-over time-scale beyond which 
the error rate $\mathcal{E}(T)$ changes its scaling from $T^{-1}$ to $\mathcal{E}(T)\sim e^4T^{-2}$.
 
In our work, we have ignored finite temperature effects despite the fact that a finite temperature is needed to justify 
the Markovian approximation. Such finite temperature effects can be expected to be negligible for temperatures substantially 
below the gap. Moreover, such finite temperature excitations are expected to only increase the excitation probability and thus 
the error rate.
In summary, we find that while a Bosonic bath does not directly interfere with topological protection of quantum information, finite (but low) temperature
Bosonic baths can lead to powerlaw in time diabatic errors that might limit the speed of topologically protected operations.

This work was supported by the  Microsoft Station Q, NSF-DMR-1555135 (CAREER), JQI-NSF-PFC (PHY-1430094) and the Sloan research fellowship.

\appendix

\section{Diabatic expansion of Bloch vector} \label{app:B}
Expanding the Bloch vector $R(s)$ in powers of $\epsilon = 1/T$,
\begin{align}
R(s) = R_0(s) + \epsilon R_1(s) + \dotso
\label{expansion1}
\end{align} 
we seek solution to 
\begin{align}
\epsilon\dot{R} = MR 
\end{align}
for initial condiction $R(0) = R_0(0) \equiv \vec{B}(0)$ where, $M = 2(A + S)$ and $S = (\alpha - \beta)A^2$ ($A$ being the matrix representation of $\vec{B}\times$ operation) with $\alpha$ and $\beta$ being the time-dependent functions defined by Eq.~ \ref{alphabetadef}. The presentation here follows the work of Hagedorn \textit{et. al.} in Ref.~\cite{Hagedorn}. 

$A$, being the anti-symmetric matrix representation of $\vec{B}\times$ operator has eigenvalues $0,i,-i$. Consequently, the instantaneous eigenvalues of $M$ are given by $0,\lambda_1 \mbox{and} \lambda_2$ with,
\begin{align}
\lambda_1 &= 2(i-(\alpha-\beta)) \nonumber \\
\lambda_2 &= 2(-i-(\alpha-\beta)).
\label{Meigenvalues}
\end{align}
Denote $R_0(s) = -\vec{B}(s)$ and thereby, $R_0(s)$ is the zero eigenvector with $0$ eigenvalue. $M$ can be inverted in the eigenvector subspace with non-zero eigenvalues,
\begin{align}
M^{-1} = \frac{1}{2(1+(\alpha-\beta)^2)}(-A-(\alpha-\beta)\mathbb{1}).
\end{align}

Expanding $R = R_0 + \epsilon R_1 + \dots $ and substituting in $\epsilon\dot{R} = MR$, we get, 
\begin{align}
MR_j &=\dot{R}_{j-1} \nonumber \\
\implies R_j &= f_{j-1}R_0 + M^{-1}\dot{R}_{j-1}
\label{ap:recursive}
\end{align}
with $f_{j-1}(s)$ evaluated using the condition  $R_0^\mathrm{T}\dot{R}_j = 0$ which follows from $MR_j =\dot{R}_{j-1}$,
\begin{align}
f_{j-1}(s) = \int_0^s ds~\dot{R}_0^TM^{-1}\dot{R}_{j-1}.
\label{ap:f_j}
\end{align}

Now we show that the series expansion is well-defined in small epsilon limit. 
Consider the partial sum of the series expansion,
\begin{align}
R^N (s) = \sum_{j=0}^N \epsilon^jR_j.
\label{MLpartial}
\end{align} 
If the series expansion is well-defined, the partial sum (as defined above) must converge to the actual solution $R$. Let the actual solution $R(s) = V(s)R(0)$. Consider,

\begin{align}
\norm*{R^N(s) - R(s)} &= \norm*{R^N(s)- V(s)R(0)} \nonumber \\
&= \norm*{V(s)}\norm*{V(s)^{-1}R^N(s)-R(0)}  \nonumber \\
&= \norm*{V(s)}\norm*{\int_0^s ds' \frac{d}{ds'} V^{-1}(s')R^N(s')},
\label{step}
\end{align}
where we have used  $\dot{V}^{-1}(0) = 0$ which follows from $\dot{\vec{B}}(0) = 0$.
Now,
\begin{align}
\dot{R}^N(s) &= \sum_{j=0}^N\epsilon^j\dot{R}_j = \sum_{j=0}^N\epsilon^jMR_{j+1} \nonumber \\
\implies \epsilon\dot{R}^N(s) &= M\sum_{j=0}^N \epsilon^{j+1}R_{j+1} + MR_0 - MR_0 \nonumber \\
&= MR^{N+1}  = MR^N + \epsilon^{N+1}\dot{R}_N,
\end{align}
where we simply used the definition given in Eq.~\ref{MLpartial} and the relation given in Eq.~\ref{ap:recursive} to arrive at the second line above.
Using the above relation we get,
\begin{align}
\frac{d}{ds'} (V^{-1}(s')R^N(s')) = \dot{V}^{-1}R^N + V^{-1}\dot{R}^N &= \epsilon^N\dot{R}_N.
\end{align}

Therefore it follows from Eq.~\ref{step},
\begin{align}
\norm*{R^N(s) - R(s)} &= \norm*{V(s)}\norm*{\int^s_0 ds'\epsilon^N\dot{R}_N} \nonumber \\
&\leq \epsilon^N \norm*{V(s)}\int^s_0 ds' \norm*{\dot{R}_N}.
\label{inequality} 
\end{align}
We conclude that $R^N$ converges to the actual solution $R$  provided $\norm*{\dot{R}_N}$ and $\norm*{V(s)}$ are bounded.
We refer the reader to Ref.~\cite{Hagedorn} for the proof of boundedness of $\norm*{\dot{R}_N}$ in absence of bath.  It seems likely that a similar proof holds for boundedness of $\norm*{\dot{R}_N}$ in presence of bath. In the limiting $\alpha(s) = \beta(s) = 0$, $V(s)$ is unitary, so clearly when for $\alpha(s) >\beta(s) \forall s$, $\norm*{V(s)}$ is bounded by 1. Without speculating about $\alpha(s) <\beta(s)$ case, we restrict our following discussion to $\alpha(s) >\beta(s)$ that corresponds to $\eta_0\geq\eta$ provided $s_x\sim s_y$. 

Our goal now is to arrive at a bound for the diabatic error defined by,
\begin{align}
\mathcal{E} = \norm{R(1)-R_0(1)},
\end{align}
where $R_0(s=1) = -\vec{B}(s=1)$ is the Bloch vector that corresponds to instantaneous zero eigenvector of $M$ at $s=1$. Note as consistency check that $R(1)\rightarrow R_0(1)$ when $\epsilon = 1/T \rightarrow 0$ follows from the series expansion of $R$, Eq.~\ref{f_j}.
Since we have shown that the series expansion of $R$ is well defined, we can  replace $R$ in $\norm{R(1)-R_0(1)}$ by its corresponding series expansion. However to make progress towards arriving at a bound for $\mathcal{E}$ we need a useful result stated and proved below.

We show that if all derivative of $M$ ($k^{th}$ derivative denoted by $M^{(k)})$, $M^{(k)}(s_0) = 0 \forall ~k $ for some $s_0$ then $(R_0^{\perp})^TR_j^{(k)}=0\forall ~k,j$ where $R_0^{\perp}$ is any vector such that $(R_0^{\perp})^TR_0 = 0$. The proof follows in three steps:

\begin{itemize}
\item We first show that $(R_0^{\perp})^TR_0^{(k)}=0\forall ~k$. \\
Let $k$ be a positive integer. $MR_0 = 0 \implies (MR_0)^{(k)} = \sum_{j=0}^k \binom{k}{j}M^{(j)}R_0^{(k-j)} = MR_0^{(k)} = 0$. Therefore, it must be $(R_0^{\perp})^TR_0^{(k)}=0\forall ~k$.

\item Next we show that $[M^{-1}]^{(k)}(s_0) = 0\forall ~k$. \\
Again, let $k$ be a positive integer. $M^{-1}M = 1 \implies (M^{-1}M )^{(k)} = \sum_{j=0}^k \binom{k}{j}[M^{-1}]^{(j)}A^{(k-j)} = [M^{-1}]^{(k)} M= 0$. Therefore, it must be $[M^{-1}]^{(k)}(s_0) = 0\forall ~k$.

\item Finally we prove our original assertion, $(R_0^{\perp})^TR_j^{(k)}=0\forall ~k,j$. \\
We will prove this assertion by induction. Assume $(R_0^{\perp})^TR_{j-1}^{(k)}=0\forall ~k$.
Now,
\begin{align}
R_j &= f_{j-1}R_0 + M^{-1}\dot{R}_{j-1} \nonumber \\
\implies R_j^{(k)} &= (f_{j-1}R_0)^{(k)} + (M^{-1}\dot{R}_{j-1})^{(k)} \nonumber \\
&= M^{-1}\dot{R}_{j-1}^{(k)} = M^{-1}R_{j-1}^{(k+1)}. \nonumber
\end{align}
Using the the induction hypothesis it follows, $(R_0^{\perp})^TR_j^{(k)}=0\forall ~k,j$.

\end{itemize}

Since $M^{(k)}(s_0) = 0;\quad s_0\in\{0,1\}\forall k$  on account of $\frac{d^k}{ds^k}\vec{B}(s_0)=0; \quad s_0\in\{0,1\}\forall k$, the above result implies $R(1)||R_0(1)$ and $R(0)||R_0(0)$.
Armed with this result, we use to Eq.~\ref{inequality} to arrive at a bound for $\mathcal{E}$ for two different cases, i.e, in absence and in presence of a thermal bath.
\subsection{Absence of thermal bath}
In absence of thermal bath $M$ is anti-symmetric and consequently $V$ is unitary.  Thus, Eq.~\ref{inequality} reduces to
\begin{align}
\norm*{R^M(s) - R(s)} \leq \epsilon^M \int^s_0 ds' \norm*{\dot{R}_M} .
\label{step1}
\end{align}
Consider the expression $\norm{R(1)-R_0(1)}$, using triangle inequality we can express,
\begin{align}
\norm{R(1)-R_0(1)} &\leq \norm{R(1)-R^M(1)} + \norm{R^M(1)-R_0(1)}. \nonumber \\
\end{align}
Using $R^M(1) \mathbin{\|} R_0(1)$ on account of all derivatives of $M(s)$ vanishing at $s=1$, 
\begin{align}
\norm{R^M(1)-R_0(1)} &= 
\norm{R_0\norm{R^M(1)}-R_0(1)} \nonumber \\
&= \abs{\norm{R^M(1)}-1} \nonumber \\
&\leq \norm{R^M(1)-R(1)},
\end{align}
where crucially, we have used $\norm{R(1)}=1$ as time-evolution is unitary for anti-symmetric $M$ to go from the second to the last line on the LHS above.

Using Eq.~\ref{step1} we get, 
\begin{align}
\norm{R(1)-R_0(1)} &\leq
2\epsilon^M \int^1_0 ds' \norm*{\dot{R}_M}.
\label{ap:exponential}
\end{align}

\subsection{Presence of purely dephasing thermal bath}
Again, using triangle inequality,
\begin{align}
\norm{R(1)-R_0(1)} &\leq \norm{R(1)-R^N(1)} + \norm{R^N(1)-R_0(1)}. \nonumber \\
\end{align}
Using $R^N(1) \mathbin{\|} R_0(1)$ on account of all derivatives of $M(s)$ vanishing at $s=1$, 
\begin{align}
\norm{R^N(1)-R_0(1)} &= 
\norm{R_0(1)\norm{R^N(1)}-R_0(1)} \nonumber \\
&= \abs{\norm{R^N(1)}-1} \nonumber \\
&= |\epsilon f_0(1) + \epsilon^2 f_1(1) + \dotso | \nonumber \\
&\simeq \epsilon|f_0(1)| \nonumber \\
\implies \norm{R(1)-R_0(1)} &\simeq \epsilon|f_0(1)| .
\end{align}

\section{Asymptotic behavior}\label{app:E}
We consider the error $\mathcal{E}$ in the presence of small but finite relaxation.
To proceed it is useful to consider a matrix $V(s)$ with initial condition $V(0) = \mathbb{1}$, that satisfies 
\begin{align}
\dot{V} &=\frac{1}{\epsilon}M V,
\label{VdiffEqn}
\end{align}
where $M$ is the matrix appearing in Bloch equation (Eq.~\ref{wnoise}). 
Using $V$ we change to change to a new variable $\xi$,
\begin{align}
\xi (s)=V^{-1}(s)R(s),
\label{formal}
\end{align}
that satisfies,
\begin{align}
\epsilon \dot{\xi} = -4\beta \xi + 4\beta V^{-1}R_0.
\end{align}
The solution to this equation is written as,
\begin{align}
&\xi(s) = e^{-\frac{1}{\epsilon}\int_0^s 4\beta(s') ds'}\nonumber\\
& \left( \int_0^s \frac{4\beta(s')}{\epsilon}e^{\frac{1}{\epsilon}\int_0^{s'} 4\beta (s'') ds''}V^{-1}(s')R_0(s')ds' + R(0)\right) \nonumber \\
\label{sol}
\end{align}
which in conjunction with Eq.~\ref{formal}, formally (i.e. contingent on having a solution for $V(s)$ in Eq.~\ref{VdiffEqn}) solves the Bloch equation (Eq.~\ref{wnoise}). Note that Eq.~\ref{sol} is completely consistent with our discussion in the $\beta = 0$ setting in Eq.~\ref{sol} and using Eq.~\ref{formal} leads to the solution $R(s) = V(s)R(0)$ where using Eq.~\ref{VdiffEqn} we find that $R$ satisfies the same Bloch equation, Eq.~\ref{BEA} as used in previous section.

For finite relaxation, the first term in the parenthesis in Eq.~\ref{sol} is non-zero and hence the matrix $V(s)$ (or equivalently $V^{-1}(s)$) must be known to calculate $R(1)$. 
However, $V(s)$ satisfying Eq.~\ref{VdiffEqn} with initial condition $V(0) = \mathbb{1}$ does not lend itself to a power series expansion in $\epsilon$ parallel to Eq.~\ref{expansion}, invalidating the method used to obtain analytical solution for the Bloch vector in the previous section.
Though there is no clear way to calculate the matrix $V(s)$ by solving Eq.~\ref{VdiffEqn}, computing $\bar{R}$, the action of matrix $V(s)$ on the initial Bloch vector $R_0(0)$,
\begin{align}
\bar{R} = VR_0(0)
\label{auxil}
\end{align}
is analytically tractable.
The solution for the vector $\bar{R}$ through Eq.~\ref{BEA} allows us to make an ansatz for $V(s)$ that leads to results consistent with numerics. 

While a direct solution of $V$ in Eq.~\ref{VdiffEqn} is difficult, we observe that the introduction of a finite relaxation does not change the solution for $\bar{R}$ in Eq.~\ref{BEA} except replacing  $\alpha\rightarrow \alpha - \beta$. Therefore generalizing Eq.~\ref{R1A}  we get, $\bar{R}(1) = f(s)R_0(1)$ where  $f(s)$ is given by Eq.~\ref{fs} but unlike previous section $\beta$ is no longer assumed to be zero. Using Eq.~\ref{auxil}, this relation can be used to constrain $V(s)$ according to the relation
 \begin{align}
 V(1)R_0(0) = f(1)U_0(1)R_0(0)
 \label{finalcondition}
 \end{align}  where we have used the relation $R_0(s) = U_0(s)R_0(0)$ where $U_0$ is defined in Eq.~\ref{basic}.This motivates our anstaz
\begin{align}
V(s) \approx f(s)U_0(s)
\label{approx}
\end{align}
in Eq.~\ref{sol} which interpolates correctly, satisfying the initial condition $V(0) =\mathbb{1}$ at $s=0$ as well as Eq.~\ref{finalcondition} at $s=1$. Note that, the above approximation satisfies $V(s)\rightarrow U_0(s) $ as $T\rightarrow\infty$. 

\begin{figure}[H]
\includegraphics[width=\columnwidth]{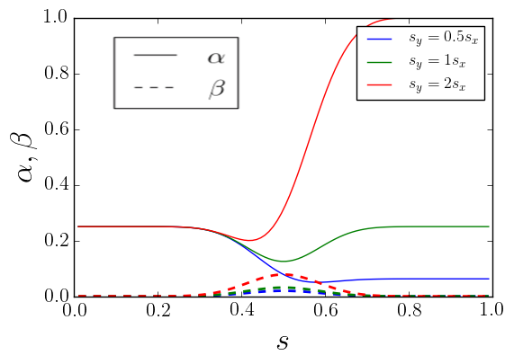}
\caption{Plot of effective dephasing ($\alpha$) and relaxation ($\beta$) as a function of scaled time, $s$ for system-bath coupling strength $s_x=0.5$ and $s_y$ given in the key. The dashed curve corresponds to effective relaxation, $\beta$ and the solid curve corresponds to effective dephasing, $\alpha$. The dephasing parameter, $\eta_0$ and the relaxation parameter $\eta$ are chosen equal to each other and set to 1.}
\label{alpha_beta}
\end{figure}

Substituting Eq.~\ref{approx} in Eq.~\ref{sol}, we make the the following ansatz,
\begin{align}
R(1) &= U_0f(1)e^{-\frac{4}{\epsilon}\int^1_0\beta}\left(\frac{4}{\epsilon}\int_0^1\beta e^{\frac{4}{\epsilon}\int^s_0\beta}f^{-1}(s)+1\right) R(0) \nonumber \\
&= \left(1 + f(1)\int^1_0 \left(\frac{d}{ds}f^{-1}\right)e^{-\frac{4}{\epsilon}\int^1_s\beta}ds\right)R_0(1).
\label{ansatz}
\end{align}
Restricting $f$ and $f^{-1}$ in the above formula to lowest order in $\epsilon$ , we get $\norm{R(1)} = 1+ \int^1_0 \left(\frac{d}{ds}f_0^{-1}\right)e^{-\frac{4}{\epsilon}\int^1_s\beta}ds$ where \begin{align}
f_0(s) &= -\int_0^s \frac{(\alpha- \beta)\omega^2}{2(1+(\alpha - \beta)^2)}. 
\label{fOrder}
\end{align}
As a consistency check note that Eq.\ref{ansatz} reduces to Eq,~\ref{R1A} in the limit $\beta\rightarrow 0$. Moreover, since $R$ must be bounded in norm $\norm{R(1)} \leq 1$,  $\frac{(\alpha- \beta)\omega^2}{2(1+(\alpha - \beta)^2)} > 0$ must hold. This condition suggests the requirement  $\alpha >\beta $ for the ansatz offered in Eq.~\ref{ansatz} to hold. We point out that $\alpha >\beta $ condition can be satisfied provided the two system-bath coupling parameters have same order of magnitude, $s_x\sim s_y$  and dephasing strength is stronger than relaxation, $\eta_0 \geq \eta$. This is clear from the Fig.~\ref{alpha_beta}, where we plot effective relaxation and effective dephasing as defined in Eq.~\ref{alphabetadef}. The dephasing parameter, $\eta_0$ and the relaxation parameter, $\eta$ are chosen as, $\eta_0=\eta = 1$.  For all the curves, $s_x = 0.5$.  We have chosen to vary just $s_y$ because both $\alpha$ and $\beta$ are invariant under combined effect of reflection about $s=0.5$ and $s_x\rightleftarrows s_y$ exchange. We see that for wide-ranging values of $s_y$, $\alpha(s) >\beta (s) \forall s \in [0,1]$. Thus we conclude, for comparable values of system-bath coupling strengths, $s_x\sim s_y$, $\eta_0>\eta$ is a good criterion to ensure $\alpha(s) >\beta (s) \forall s \in [0,1]$.

Using Eq.~\ref{ansatz}, the error may be computed as 
\begin{align}
\mathcal{E}(T) = \norm{R(1)-R_0(1)} &= f(1)\int_0^1 (\frac{d}{ds}f^{-1}) e^{-4T\int_s^1\beta}ds \nonumber \\
&\approx -\frac{1}{T} \int^1_0 \dot{f}_0  e^{-4T\int_s^1\beta}ds,
\label{fullTh1}
\end{align}
with $\dot{f}_0 = \frac{(\alpha- \beta)\omega^2}{2(1+(\alpha - \beta)^2)} $ and $\alpha$,$\beta$ being defined according to Eq.~\ref{alphabetadef}.

\begin{figure}[H]
\includegraphics[width=\columnwidth]{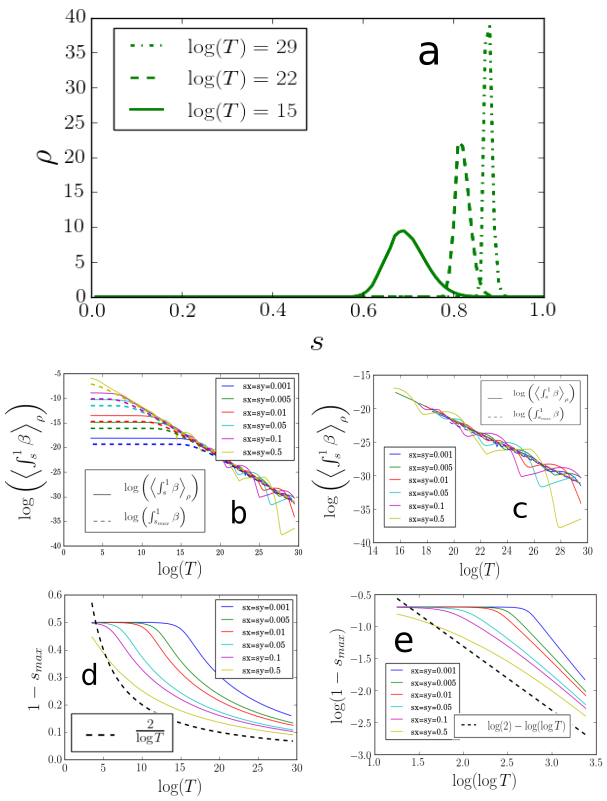}
\caption{(Panel a) Plot of $\rho$ (Eq.\ref{rho}) for different values of time T. The peak-height and the width of the probability density $\rho$ is an increasing and a decreases function of T, respectively. The probability densities look identical at this scale for values of $s_x=s_y$ varied from 0.005 to 0.5 (i.e. over two orders of magnitude).
(Panels b and c) Comparison of $\left\langle\int_s^1 \beta\right\rangle_\rho$ (defined by Eq.~\ref{def_int_beta_av}) , plotted in solid curves, versus its approximate estimate given by values of $\int_{s_{\textit{max}}}^1 \beta$ as a function of total time $T$ (in the units of $1/\Delta$), plotted in dashed curves, for different values of dephasing, $\alpha$ and relaxation strength, $\beta$ characterized by the shown values of $s_x = s_y$ (see Eq.~\ref{wnoise}-\ref{F}). Relaxation and dephasing parameters, $\eta$ and  $\eta_{0}$, that solely depend on bath Hamiltonian are both set to $\eta=\eta_{0}=1$. Both sets of curves, solid and dashed, are computed numerically. There exists a region, for small values of $T$, over which $T$ dependence of $\left\langle\int_s^1 \beta\right\rangle_\rho$ vanishes (panel c). This region is well captured by the approximate formula $\int_{s_{\textit{max}}}^1 \beta$, however the value of $\left\langle\int_s^1 \beta\right\rangle_\rho$ itself is underestimated by the formula. For large values of $T$, $\left\langle\int_s^1 \beta\right\rangle_\rho$ values tend to oscillate, however, the average slope is again well captured by the approximate formula $\int_{s_{\textit{max}}}^1 \beta$ (top left panel). \\
(Panels d and e) Plot of $1-s_\textit{max}$ as a function of $\log T$ (bottom right panel) and the same plot in log-log scale is shown in bottom left panel. Notice that $T$ independent region of $\left\langle\int_s^1 \beta\right\rangle_\rho$ corresponds to $s_\textit{max}\approx 0.5$. $s_\textit{max}\approx 0.5$ region ends in a kink beyond which $1-s_\textit{max}$ decreases zero, asymptotically approaching $\frac{1}{\log T}$. This asymptotic dependence is verxy the log-log plot in the bottom right panel.}
\label{justification}
\end{figure}

Now, we estimate the exponent of $T$ which governs the power-law  dependence of $\norm{R(1)-R_0(1)} = \mathcal{E}$ for large $T$. For our convenience we will restrict ourselves to the special case $s_x = s_y$, (see Eq.~\ref{wnoise}-\ref{F}) essentially allowing the system bath coupling to be governed by a single parameter. 
Taking the derivative of $\mathrm{log}(\mathcal{E})$with respect to $T$, 
\begin{align}
\frac{d}{dT}\mathrm{log}(\mathcal{E})\approx -\frac{1}{T} - \frac{4\int_0^1 ds \dot{f}_0e^{-4T\int_s^1\beta}\int_s^1\beta }{\int_0^1 ds \dot{f}_0 e^{-4T\int_s^1\beta} }.
\end{align}
Defining,
\begin{align}
\rho \equiv \frac{\dot{f}_0 e^{-4T\int_s^1\beta}}{\int_0^1\dot{f}_0 e^{-4T\int_s^1\beta}}
\label{rho}
\end{align}
as a probability density defined over $[0,1]$, the second term in the equation above is interpreted as 
\begin{align}
\left\langle 4\int_s^1\beta\right\rangle_\rho \equiv \frac{4\int_0^1 ds \dot{f}_0e^{-4T\int_s^1\beta}\int_s^1\beta }{\int_0^1 ds \dot{f}_0 e^{-4T\int_s^1\beta} }.
\label{def_int_beta_av}
\end{align}
The function $\dot{f}_0 (s)= -\frac{(\alpha-\beta)\dot{\theta}^2}{2(1+(\alpha-\beta)^2)}$ (Eq.~\ref{fOrder}),  is symmetric about $s=0.5$ (when $s_x = s_y$) and exponentially goes to zero at $s=1$, while, the function $e^{-4T\int_s^1\beta}$ is an increasing function over $[0,1]$ where, it decreases exponentially from the value $1$ at $s=1$ to the value $e^{-4T\int_0^1\beta}$ at $s=0$ with exponent being proportional to $T$. These properties imply that $\rho$ is a sharply peaked (see Fig.~\ref{justification}a) distribution for large $T$ (defining large $T$ when $1/T \ll \int_0^1\beta$ holds) with the maximum value $\rho_{\textit{max}} = \rho(s_{\textit{max}})$ for $s_{\textit{max}} \in \big(0.5,1\big)$. Moreover, $s_{\textit{max}} \rightarrow 1$ as $T\rightarrow \infty$. Hence for large value of $T$, we approximate (see Fig.~\ref{justification}),
\begin{align}
\left\langle \int_s^1\beta\right\rangle_\rho  \simeq \int_{s_{\textit{max}}}^1\beta.
\label{betaexpectation}
\end{align}

$s_{\textit{max}}$ is the solution to the equation $\rho '(s)|
_{s_\textit{max}} = 0$,
\begin{align}
0 = \left.\frac{\dot{\Omega}}{\Omega}   + 2\frac{\dot{\omega}}{\omega} + 4T\beta \right|_{s_{\textit{max}}}
\label{smax}
\end{align}
where, we defined $\Omega \equiv\frac{(\alpha-\beta)}{2(1+(\alpha-\beta)^2)} $ and $\omega = \dot{\theta}$ (see Eq.~\ref{theta}) for brevity.
Since $\Omega(s \rightarrow 1) \neq 0$ and $\Omega ' (s\rightarrow 1) = 0$, we conclude  $\left|\frac{\dot{\Omega}}{\Omega}\right| \ll \left|\frac{\dot{\omega}}{\omega} = -\frac{2s-1}{s^2(1-s)^2}\right|$ for $s\rightarrow 1$. Thus, neglecting  $\frac{\Omega '}{\Omega}$  term in the Eq.~\ref{smax}, the equation for $s_{\textit{max}}$ in the $T\rightarrow \infty$ limit is given by 
\begin{align}
\frac{2s_{\textit{max}}-1}{4T} \simeq \frac{1}{2}s_{\textit{max}}^2(1-s_{\textit{max}})^2\beta(s_{\textit{max}}).
\label{s_maxeq}
\end{align}
This result immediately leads to two conclusions. First, using the asymptotic form of $\beta$,
\begin{align}
\beta \overset{s\rightarrow 1}{\simeq} \eta\left(\frac{\pi}{4C}\right)^2 (s_x^2 + s_y^2)s^4(1-s)^4 e^{-\frac{2}{1-s}},
\end{align}
where $C \equiv \int_0^1 ds' e^{-1/s'(1-s')}$, one finds
$\mathrm{log}(T) \sim \frac{2}{1-s_\textit{max}} + O(\mathrm{log}(1-s_\textit{max}))$, and thus,
\begin{align}
s_{\textit{max}} \sim 1-\frac{2}{\mathrm{log}(T)}.
\label{smaxsol}
\end{align}
Second, using asymptotic dependence of $\int_s^1\beta$ on $s$ given by,
\begin{align}
\int_s^1\beta ds \overset{s\rightarrow 1}{\simeq} \frac{1}{2}s^2(1-s)^2\beta(s). 
\end{align}
Taken together with Eq.~\ref{s_maxeq}, we get
\begin{align}
\frac{2s_\textit{max}-1}{4T} \simeq \int_{s_\textit{max}}^1\beta ds.
\end{align}
Thus,
\begin{align}
\frac{1}{\mathcal{E}}\frac{d\mathcal{E}}{dT} &\approx -\frac{1}{T} - \left\langle 4\int_s^1\beta ds\right\rangle \nonumber \\
&\simeq -\frac{1}{T} -  4\int_{s_\textit{max}}^1\beta ds\nonumber \\
&\simeq -\frac{2s_\textit{max}}{T}.
\end{align}

In conclusion, $\norm{R(1) - R_0(1)}$ is given by 
\begin{align}
\norm{R(1) - R_0(1)} \sim T^{-2s_\textit{max}}.
\label{asymptotic_form}
\end{align}
Defining the exponent at time $T$ as 
\begin{align}
m(T) \equiv -2s_\textit{max} \sim -2 + \frac{4}{\log T},
\label{m}
\end{align}
where we have used the asymptotic dependence of $s_\textit{max}$ on $T$ (see Eq.~\ref{smaxsol}).
Since $s_\textit{max}(T) \in (0.5,1)$, the exponent $m(T) \in (1,2)$ with $m(T\rightarrow \infty) = 2$. All assumptions leading upto this result are verified against exact numerical results in Fig.~\ref{justification}.

\bibliography{ref}
\end{document}